\documentclass[aip,jcp,reprint,dvips]{revtex4-1}

\begin{document}

\title{Autobiography of Yoshitaka Tanimura}\thanks{\href{https://doi.org/10.1021/acs.jpcb.1c08552}{https://doi.org/10.1021/acs.jpcb.1c08552}}

\author{Yoshitaka Tanimura}
\email{tanimura.yoshitaka.5w@kyoto-u.jp}
\affiliation{Department of Chemistry, Graduate School of Science, Kyoto University, Kyoto 606-8502, Japan}
\date{\today}

\maketitle

\section*{Prologue}

My name is Yoshitaka. {\it Yoshi} means ``fortune," and {\it Taka} means ``rising." When I was a child, I did not like my name because it was not very common. I was born in Kobe and grew up in Tokyo. My father was a jet-engine engineer at Kawasaki; he was an overworked Japanese businessman. I went to public schools, and my grades were quite average. I was just a normal boy, except that I enjoyed reading science books, on such topics as the theory of relativity and quantum mechanics. My sister was studious, while my brother was a technology geek.  He imported a personal computer directly from the U.S, which was probably one of the first to arrive in Japan. He is now the president of a video game company. 

I ceased being a normal boy when I took up rock climbing and ice climbing in high school. At the age of 17, I was trapped on a steep, wintery mountain with friends for a few days. There was quite a fuss about the incident on TV and in the newspapers. I got frostbite on all of my toes. 

Although I was good at math and physics, I detested studying. For this reason, I was not good at English, Biology, Chemistry, or any other subject that required diligence. While I did not study much, I was nonetheless able to enter the Faculty of Engineering at Keio University, which is one of the best private universities in Japan.

\section*{Keio University}

During my time as a freshman, my commitment to mountain climbing increased substantially. I spent more than 200 days per year climbing mountains. When I was a sophomore, I traveled to the Indian Himalayas and reached the summit of Mt. Kharuchakund (6632m) in the Gangotris, where the Ganges river goddess is said to have touched the Earth for the first time. I aspired to become a professional mountaineer and climb each one of the 8000m-class mountains. Meanwhile, my grades in the university were terrible.

When I was 20, my friend died in a rock-climbing accident. This event changed my life completely. I gained a strong desire to understand the purpose and meaning of life, with a similar emotion that had previously driven me to the mountains. At first, I was interested in philosophy, but 
after noticing that my friends in the department of philosophy did not do anything philosophical, my interest died. I then came to the conclusion that human beings have not yet evolved enough to answer such deep questions as why we are here, and that humans are still in the process of laying the foundation for such knowledge. In the end, I became convinced that the only way to reach the truth was through basic science, a subject in which I had always been interested. I knew, however, that there was no way that I could 
lay down the foundation for such knowledge all by myself. Still, I hoped that through basic science, I could help future generations reach the ``truth."

The idea of conducting basic research in theoretical physics was a hopelessly high summit for a student with such terrible grades. In fact, until that time, I rarely went to the university campus. As a consequence, I needed to repeat my sophomore year. Moreover, because I was in the engineering department, I could not learn physics systematically in the classroom. For this reason, I needed to study the basics of physics on my own. I did this by taking advantage of my extra sophomore year. In particular, I carefully studied the Landau and Lifshitz series. I also studied math whenever I needed to. This was the period of my life during which I studied the most intensively. By the fall of my junior year, I had mastered mechanics, quantum mechanics, and general relativity. However, by the time that I thought I had reached my goal of mastering physics, I was devastated to realize that there is a world of difference between learning science and conducting scientific research.

\subsection*{Ryogo Kubo}

Right around that time, I was informed that Keio University was going to establish physics and chemistry departments and reorganize them into the Faculty of Science and Engineering. I also learned that Keio University was home to a world-renowned physicist, Ryogo Kubo. 

Thanks to the year-long delay in my studies due to repeating my second sophomore year, I thought that I might be able to join the Kubo group. One time, while I was talking to a friend about this possibility near the university cafeteria, I noticed an old gentleman who was looking at me from a short distance away. A month later, there was an announcement that Professor Kubo would be accepting students from other departments. I rushed to visit his office and saw that old gentleman sitting there. That was my first contact with Kubo-sensei. 

I joined the Kubo Group in my senior year, along with other students from other departments. Several members of the theoretical physics faculty took turns teaching us how to read important scientific papers, such as by Feynman, Bloch, Landau, and Kubo. I worked like a dog. 

\subsection*{Stochastic theory}

During the second half of my senior year, I conducted my research project under the guidance of Kubo-sensei. The theme of my thesis was the analysis of the nonlinear optical spectra of randomly modulated multi-level systems, as described by Kubo's stochastic Liouville equation (SLE). While this approach is phenomenological, it facilitates analytic calculations of third-order optical response functions of systems under non-Markovian and non-perturbative fluctuations. The capabilities of this approach were apparent: whereas the Markovian quantum master equation (QME) approach predicts only a Dirac delta-shaped Raman peak and a Lorentz-shaped phosphorescent peak, the SLE approach predicts these two peaks and also a broadened Raman peak of Lorentzian form that appears at the Raman peak position. 

After struggling for two months, I obtained, with the help of Hiroshi Takano, a research associate in the Kubo group, an explicit analytic expression for the broadened Raman peak. I then discovered that this peak has peculiar properties, such as being a part of the Raman spectrum in the fast-modulation limit and being a part of the phosphorescence spectrum in the slow-modulation limit.

When I proudly explained my results to Kubo-sensei, he just listened quietly. What surprised me was that a week later, he developed a general theory based on my modest results and derived my expression as an application of it. It reminded me of the saying, ``A frog in a well does not know the ocean." I based my senior thesis on these results.

Kubo-sensei was a quiet and thoughtful person. For a long time after I met him, I had no idea what he was thinking. His instructions were ambiguous and hard to understand. He was my Yoda. 

After entering graduate school, I rewrote my thesis in English in order to submit it to J. Phys. Soc. Jpn. (JPSJ). This punishing process took a full year, due to the fact that I had not been diligent in my studies of English. Even then, the quality was low. Around that time, Kubo-sensei became the president of the Science Council of Japan and was very busy. In total, it took two and half years for my senior thesis to be published (1; henceforth numbers in parentheses denote relevant papers in Publication List).
 
While waiting for my paper to be completed, I began mountain climbing again. To my great surprise, my rock climbing skills had improved, even though I 
had  been away from a rock wall for over three years. Although I did not climb nearly as often as I had previously, the routes that I climbed became extremely technical.

\subsection*{Hierarchical equations of motion (HEOM)}

The broadened Raman process that we studied was obtained from the phenomenological SLE. The noise described by the SLE is non-perturbative and non-Markovian, but, at that time, there was no Hamiltonian-based dynamical theory that described such phenomena. For this reason, I was not sure if the process really existed. The SLE can handle such noise because its Fourier-Laplace-transformed solution can be expressed in a continued fraction form with distinctive convergence. I wondered if I could find an equation of motion (EOM) based on a Hamiltonian system with a form similar to that of the SLE.

Around that time, several master course students joined the Kubo group as my juniors. We studied the Feynman-Vernon theory, in which the time evolution of a system that interacts with a bath is described by the reduced density operator (RDO) in a path integral representation. The effects of the heat bath are incorporated into the RDO as the influence functional in terms of dissipation and temperature-dependent fluctuations characterized by the spectral distribution function (SDF) of the system-bath (SB) interaction. They  are related through Kubo's fluctuation-dissipation theorem (FDT), and the total system approaches the thermal equilibrium state in the long time limit. 

I suspected that the fluctuation in this theory corresponds to the random noise in the SLE. As a first step in exploring this point, I differentiated the RDO with respect to time to obtain the EOM. I reported my finding in a group meeting. However, Kubo-sensei was very sceptical of my approach and angry, commenting, ``Differentiating a density matrix expressed in terms of a functional integral does not provide anything new." I was very disappointed. But after a few days, I realized that a path integral is an integral representation in configuration space, and for this reason, it cannot be undone by differentiating it with respect to time. 

I brought this point up at the next group meeting, and I took the discussion a step further by presenting the derived QME. Kubo-sensei listened with a grim face, and he replied ``The TLS is expressed in terms of energy states and not in configuration space. This system cannot be expressed in path integral form." However, the following week I showed that the RDO for the TLS could be expressed in terms of path integrals if one employs the coherent state representation in Grassmann variables. Around that time, I found that Caldeira and Leggett had derived the quantum Fokker-Planck equation (QFPE) for a Markovian bath from the Brownian Hamiltonian using an approach similar to mine. 
Gradually, Kubo-sensei came to approve of my method.

Assuming that the SDF is Ohmic and that the bath temperature is high, both the fluctuation and dissipation terms take delta-function forms (Markovian condition), and the QME is obtained, as in the case of the QFPE. In contrast to the conventional QME, this equation has a fluctuation term that is proportional to the temperature. The steady-state solution of this QME is the thermal equilibrium state at finite temperature. While this equation is applicable only at high temperatures, the rotational wave approximation is not required.

From this QME, I realized that if I replaced the Ohmic SDF with the Drude SDF, the fluctuation in the QME would exhibit the same time correlation as the random noise in the SLE. I knew that the SLE can be rewritten as a simultaneous differential equation expressed in terms of the system density operators in the stochastic eigenstates. With reference to these equations, I differentiated the RDO in the Drude SDF case. Instead of using a perturbative approximation, I replaced the term describing the relaxation with an auxiliary density operator (ADO). When I differentiated this ADO, a new ADO appeared, in addition to the terms of the original RDO and ADO. By repeating the process of time differentiation for the ADOs, I obtained a hierarchy of the EOM, as in the case of the SLE. What I obtained is a prototype of the EOM which would later be called the hierarchical equations of motion (HEOM). 

The HEOM are constructed in the framework of a dynamical theory: unlike the SLE, the HEOM contain not only fluctuation but also dissipation. Because of the presence of the dissipation, which is related to the fluctuations through the FDT, the steady-state solution of the HEOM is a thermal equilibrium state at finite temperature. Best of all, the HEOM constitute a non-perturbative, non-Markovian dynamical theory. These results were obtained in 1986; however, Kubo-sensei was so busy that he could not read my paper until 1988. 

By this time, there were only myself and Takano-san in the Kubo group. While Kubo-sensei treated me like a grandson, I had to report something at every group meeting. Because of this pressure, I extended the HEOM method to treat nonlinear response functions of physical observables and wrote three papers. There were times when not only Kubo-sensei but also Takano-san would fall asleep, and I would have to wait for about half an hour for them to wake up. 

In the middle of 1988, Kubo-sensei finally realized that my doctoral term was nearing its end. So he began to proofread my manuscripts. The EOM that I derived were an extension of the master equation, and because I derived them during my doctoral course, I named them the ``doctor equations." Unfortunately, this part of manuscript was deleted by Kubo-sensei.

I wished to submit my papers to Physical Review because I wanted to work in the U.S. after getting my Ph.D. However, Kubo-sensei emphasized the importance of maintaining the scientific identity of Japanese journals, and for this reason, he only allowed me to submit my papers to JPSJ (2). While Kubo-sensei had taken a long time to read this paper, he immediately proofread my other papers when I told him that I would also submit the other three papers to JPSJ (3-5). These four papers formed the basis of my doctoral dissertation.

After each of these papers had been accepted, I wanted to get Kubo-sensei to admit that my derivation method was not worthless. I asked him, ``Kubo-sensei, you said that the time differentiation of the RDO expressed in terms of path integrals would not produce anything new, but it did, didn't it?" In response, he said, ``Did I say that?!" He was always my master. 

To broaden my horizons and to improve my English ability, I decided to go abroad after my doctorate. However, before I left Japan, I went free solo climbing on the rock wall where my friend had died eight years before. It took me a very long time to go there, but once I started climbing, I finished within only 30 min, a tenth of the normal time.

\section*{The University of Illinois at Urbana-Champaign (UIUC)}

After my thesis defense, I contacted Prof. Tony Leggett and asked if there were any postdoctoral positions available. This process took some time because we were communicating by airmail, but in the end, Tony arranged a shared postdoc position for me between himself, Prof. Peter Wolynes in  Chemistry, and Prof. Karl Hess in Electrical Engineering. I was not sure if I could work on chemistry problems, because my grades in chemistry had been poor. However, in the end, I decided to go to the UIUC because I felt that it would be a good opportunity to broaden my horizons.

In May 1989, I started new research projects at Beckman Institute, which looked like a five-star hotel. New horizons were indeed abound at the UIUC.

\subsection*{Low-temperature corrected HEOM}

With Karl, I worked on investigating the thermal effects of electron conductivity in  nanowires. As a first step toward investigating phonon baths, I extended the HEOM to treat arbitrary SDFs and temperatures using the Fourier representation of the hierarchy. As an application, I introduced low-temperature correction terms. The most commonly used form of the HEOM are described in this paper (6), but the results were buried for almost 15 years because they could not be verified as even a high-performance computer at that the time had only 32 MB of memory. I honestly did not believe that the necessary computations would be possible in my lifetime.

\subsection*{Hierarchical quantum Fokker-Planck equations (HQFPE)}

With Peter, I had initially planned to work on a protein folding problem, but we ended up working on chemical reaction (CR) problems described by a SB model instead. In the past, Peter had used the method based on the variational principle to study quantum effects of CR processes. He proposed extending the HEOM to apply it to systems with double-well  potential energy surfaces (PES). This problem is also closely related to the application of the Caldeira-Leggett theory to the quantum measurement problem, which was an attempt to study the foundations of quantum mechanics.

For this system, I extended the HEOM to employ the Wigner distribution function (WDF) describing the configuration space. Up to this point, both the SLE and the HEOM had only been solved by using the Laplace transformation in a continued fractional form. However, the WDF space is too large for the Laplace transformation, so I tried to solve an infinite series of hierarchical differential equations directly. 
Then, I discovered that there is a simple relationship between the higher hierarchical ADOs that could be used to terminate the series of equations without a loss of accuracy. This result was a milestone in the development of the HEOM (7).

By numerically integrating the hierarchical quantum Fokker-Planck equations (HQFPE), we can treat PESs of any shape. Moreover, we can apply time-dependent external fields of any profile and any strength to the system at no extra computational cost. The HQFPE are an extension of the Markovian QFPE. In the classical limit, this set of equations become an extension of the Kramers equation to the case of non-Markovian noise. The existence of the classical counterpart makes it possible to easily identify quantum effects by comparing the results of the quantum and classical calculations under the same physical conditions. Using the HQFPE, we calculated the CR rate in the quantum and classical regimes as a function of the bath coupling strength, noise correlation, and temperature, and we found that the quantum results also exhibit Kramers turnover behavior (8).

I thought that I had give up mountain climbing after moving to Illinois, where the horizon was wide open. However, a friend told me that if I drove 1,000 miles west, I could reach the Rocky Mountains. Thus I made several trips there. I enjoyed not only mountain climbing but also driving my car (a Pontiac Fierro). During the Christmas period, I drove 8,000 miles in 10 days through Louisiana, Texas, Arizona, and several other sates, and hiked in the mountains. Eventually, I decided to climb to the highest point in each of the 50 states. 

At that time in Japan, the majority of researchers in statistical physics were working on critical phenomena in spin-lattice systems, while I was working on open quantum dynamics theory. Through interactions with Peter, Karl, Tony, and their postdocs and students, I realized that the research that I was conducting was a major field in the U.S. In particular, my field is well aligned with chemistry.
The most advanced experiments in statistical physics were conducted in physical chemistry laboratories.

I was at the UIUC for a little over two years. After I completed some of my research there, I applied to several universities in the U.S. for a faculty position with the hopes of further broadening my horizons. However, the U.S. was in a recession then, and positions were scarce. To make matters worse, the Soviet Union was collapsing, and the big-name Russian scientists were coming to the U.S. It seemed impossible for me to get a faculty position. Thus, I contacted Prof. Shaul Mukamel in the Department of Chemistry at the University of Rochester and inquired about whether there were any postdoctoral positions available. He promptly replied, telling me that I should come to Rochester immediately. I moved to Rochester in the fall of 1991, feeling guilty for not being able to do any more work for Karl or Tony.

\section*{University of Rochester}

I thought that Illinois was a cold place.  However, after moving to Rochester, I realized that Illinois was not so bad. That winter, the temperature in Rochester dropped to -32 degrees Celsius. 

Shaul was fully conversant with the trends in ultrafast nonlinear spectroscopic experiments and had gathered postdocs from a variety of backgrounds to do research that would help experimentalists by pointing the way to observations of novel phenomena. He was diligent and came to each of the five postdocs twice a day for intense discussions. We called these the ``morning and evening services."

\subsection*{Multi-state quantum Fokker-Planck equation (MSQFPE)}
For my first project at Rochester, I extended the QFPE to treat electronic transitions with arbitrary forms and time dependences of non-adiabatic and laser interactions among electronic PESs (MSQFPE). Then, I applied this method to calculate ultrafast nonlinear spectra. 
At that time, Shaul was trying to develop a method to calculate the third-order response function by splitting it into a laser excitation part and a probe part (the doorway-windows picture). However, this method did not work with a system coupled to a heat bath. He asked me to clarify the reason for this. So I said I would do it when the MSQFEP project was completed, but then he began coming to my office every day saying, ``Taka, you must do it!" At last, after a week, he said, ``Taka, please...," so I had to suspend the MSQFEP project to carry out the tedious path integral calculations.

The long, quiet, snowy winter nights of Rochester were perfect for doing such calculations. By extending the analytic solution of the RDO for Brownian systems obtained by Herman Grabart and his colleagues, I derived the generating functional (GF) that could be used to calculate nonlinear response functions of arbitrary order (9). Then, I found that the reason that the doorway-windows picture did not work was because it ignores the quantum coherence between the system and bath. Without this coherence, for example, there would be no echo peak in the photon echo measurement, for example.  I then realized that in the HEOM formalism, this coherence is incorporated into the ADOs, and hence that the HEOM could be used to treat time-dependent external fields and calculate nonlinear response functions.

\subsection*{Multi-dimensional Spectroscopy}
When I had almost completed this work and was ready to get back to the MSQFPE project, Shaul came to my office and suggested calculating the seventh-order Raman echo signal for a Brownian system (which had been used as a model for molecular liquids) with the GF approach. I did this calculation and found that not only was there no echo, but that not even a signal was generated in this model. Shaul was upset with my results, but that's just how the math worked out! However, by including a nonlinear component of the polarizability, I constructed a model that did generate a signal, not only at seventh-order, but even at fifth-order. Because these results did not exhibit clear echo peaks, I plotted two-dimensional (2D) spectra as functions of the time interval between the excitation and detection pulses. These 2D profiles are turned out to be much more sensitive to environmental conditions than those in conventional spectroscopy.

Inspired by 2D NMR, Shaul had been thinking about the possibility of 2D laser spectroscopy. He was thus pleased with the 2D Raman profiles that I plotted. We pointed out that similar experiments could be done with infrared (IR) lasers, although the light sources were limited at that time. Together with the expressions for electronically resonant measurements, we published these results as the theory of the fifth-order 2D Raman spectroscopy (11).  I had no idea, however, that this kind of measurement would later become a very common reality in ultrafast coherent 2D spectroscopy studies.

After this 2D Raman paper was submitted, Shaul came to my office and asked me with full smile, ``Taka, somehow you stopped doing the MSQFPE project. Why is that?" He had finally given me permission to finish the MSQFPE investigations (14). 

I was in Rochester for two and a half years. I spent weekends driving to visit the highest points in the 20 eastern states. Guanha Chen and Vladimir Chernyak admired the fact that somehow I was able to accomplish this even while being under Shaul's strict labor management.  

After spending five years in the U.S., I felt that my research would develop better if I was in a chemistry department, where intensive discussions between experimentalists and theorists are common. Thus I applied for a faculty position at the Institute for Molecular Science (IMS) in Japan. With strong support from Shaul and Dwayne Miller, I was appointed as an associate professor there. Shaul was as happy as if he had been me.

\section*{Institute for Molecular Science (IMS)}

I moved to IMS in April 1994, at the age of 33. Unlike at Japanese universities, associate professors at the IMS were independent, well-funded, and had their own research associates and postdocs. While there were Ph.D. students, the teaching duties were negligible. Hiroki Nakamura, who was the head of the theoretical division, said that if he was reincarnated he wished to return as an associate professor at the IMS. 

Ko Okumura joined my group as a research associate and we worked on developing analytic theories for nonlinear response functions, including those for 2D Raman spectroscopy. Yoko Suzuki and Tsuyoshi Kato also worked on various 2D spectroscopy theories using various approaches. 

I supervised three students during the doctoral program. Whenever students came to me for a position in my group, I told them the truth: "Theoretical research is difficult, life is short, and most importantly, there are few jobs." Moreover, as is still the case with many Japanese universities, the students had to pay their tuition and living expenses themselves. With this in mind, I told them that if they do not mind having the miserable lifestyle which goes with the pursuit of basic scientific research, they are welcome. Without flinching at those words, they joined my group. Since then, and even after I moved to Kyoto University, I have been giving the same words of advice to students who wish to join my group. I found this to be the most efficient way to identify those students with high aspirations for doing basic science. 

IMS students entered the graduate program from other universities after obtaining a master degree. I took advantage of their various backgrounds to expand the scope of my research. With the help of the computer-oriented student, Yutaka Maruyama, the MSQFPE was extended to compute pump-probe spectra in the non-Markovian case (22). With the help of quantum chemistry-oriented student, Yutaka Imamura, a physical model for organic superconductor materials was developed (39). Efficient methodologies for calculating correlated electronic states were also investigated with Osamu Hino (50). I had no knowledge of quantum chemistry, but Seiichiro Ten-no, who has a broad knowledge of science and is one of the best quantum chemists in Japan, helped us to navigate from beginning to end. He has been one of my closest friends since that time.

The IMS was Japan's gateway for foreign researchers in physical chemistry  at that time. Mitsuo Ito, the director of the IMS, encouraged me several times to organize international workshops, conferences, and schools. Through these activities, I met many experimentalists, including Graham Fleming, Andrei Tokmakoff, Thomas Elsaesser, and Peter Hamm. I realized that the best experimentalists are also theorists who could analyze their results using a simple model without calculations. I also hosted Biman Bagchi, Minhaeng Cho, Jose Onuchic, Oliver Khün, and Eok-Kyun Lee as visiting professors. I climbed Mt. Fuji with Onuchics, and later with Lees. 
It was at the IMS where I also got to know Keisuke Tominaga, Tahei Tahara, Shinji Saito, Katsuyuki Nobusada, and Iwao Ohmine.

After several years of research at the IMS, I realized that I was not taking full advantage of being there. This prompted me to visit Mont Blanc, Denali, and several other mountains. I climbed Mt. Aconcagua (6961m) with Ryo Akiyama and Hitoshi Sumi.
I had a privileged research life at the IMS, but my position was reserved for young researchers. Moreover, I wanted to pass on my experiences and knowledge to students who were as passionate about basic science as I was. So in 2003, I left the IMS and became a professor in the Department of Chemistry, Faculty of Science at Kyoto University, at the age of 43.

\section*{Kyoto University}

What makes the Faculty of Science at Kyoto University unique is its commitment to basic science. There is a tradition there of professors conducting ``curiosity-driven" research, without regard for the results. For example, there was a professor who kept fruit flies in the dark for 50 years (3000 generations)  to see what had changed.  I taught a yearlong statistical thermodynamics class for undergraduate chemistry students, but at Kyoto University, there was no syllabus nor class registration. So I taught advanced statistical mechanics, including stochastic theory and Kubo's linear response theory.

Soon after I arrived, several students joined my group from inside and outside of the university, even though I had warned them that they would have a miserable life. Among them were Akihito Ishizaki, Yuki Nagata, Yoichi Suzuki, and Taisuke Hasegawa. Most of these students were conducting research while working part-time outside the university to support themselves. However, they were highly motivated and diligent students who were my comrades-in-arms. 

By that time, my research consisted of two pillars, the HEOM and multidimensional spectroscopy. Although these involve different types of methodologies, they both have a wide range of applications. Aki made advances through which the HEOM could be used to treat low-temperature systems (62), and Yuki and Taisuke developed methods to compute 2D spectra directly from molecular dynamics (MD) simulations (64,67,75), by extending the works of Shinji and Thomas Jansen. Our work with MD simulations was initiated because the 2D Raman measurements were so difficult that experimentalists no longer paid attention to our theoretical predictions unless we estimated the signal strength. These HEOM and MD studies provided the basis for further developments of our research program.

Shortly after moving to Kyoto University, I married Ayumi, a graduate student of Fumio Hirata, at Tokyo Disneyland Resort. My parents and friends were shocked and congratulated us. We had our first and second sons, Ryota and Koichi, a little later, and I realized that my kids were a hundred times cuter than my graduate students, whom I had always thought of as my own children. 

Undergraduates in chemistry and physics also began to join my group regularly in search of their miserable life. 
Most of the graduate students were not in a hurry to acquire results and enjoyed spending their time working on basic science. 

Yumiko Ueno, our group's incredible secretary, was devoted to supporting group members and foreign visitors alike. She had a profound influence on all of our lives. We were deeply saddened when she passed away suddenly from cancer last summer.

\subsection*{The HEOM}
After the success of Aki and Graham with their application to energy transfer, the HEOM, which had been completely unnoticed for nearly 20 years, had gradually gained recognition. Thanks to the contributions of excellent researchers, including Yijing Yan, Qiang Shi, Jianshu Cao, and Michael Thoss, and many other researchers who are also my friends, and thanks to advances in computing technology, the range of systems that the HEOM can treat have become considerably broader (130). Together with Atsunori Sakurai, Akihito Kato, and others, I re-examined well-known problems, such as those of resonant tunneling diodes (RTD), quantum ratchets, and electron transfer (ET), which had only been investigated using the Markov approximation, which is not applicable to low--temperature systems. We found various peculiar effects, such as self--excited current oscillation in RTD (99,102,112) and the suppression of ratchet current by quantum tunneling (100). Daniel Packwood, a postdoc from the New Zealand, explored a non-Gaussian stochastic process (89). When Midori Tanaka was a master-course student, she derived the HEOM for a system of displaced Brownian oscillators at low temperature (82, 85). I named these the ``real" master equations.

\subsection{2D spectroscopies}
The HEOM approach is characterized by its capability to treat low-temperature, non-perturbative, non-Markovian SB interactions, and it is ideal for computing 2D spectra. In general, the microscopic Hamiltonian-based MD approach is bottom--up, whereas the model-based HEOM approach is top--down. These two approaches are complementary: the MD approach is helpful for analyzing the microscopic motion of molecules and also allows us to determine the parameters of the SB model (106), while the HEOM approach is helpful for obtaining macroscopic information regarding collective molecular motion and also allows us to carry out quantum simulations of 2D measurements (87), which are not feasible with the MD approach.  

To facilitate our research, Taisuke developed a water potential for the spectral analysis of intermolecular and intramolecular vibrations (88). Nobuhiro Ito and Ju-Yeon Jo then simulated and analyzed 2D THz--Raman (104, 116), 2D IR--Raman (113), and single-beam 2D Raman spectra (134). Through the MD investigations, we collaborated with experimentalists, including Dwayne (79) and Peter Hamm, although Peter now may be better described as a theorist.

As the applicability of the HEOM became broader, my curiosity increased. At the same time, there was an increase in the number of graduate students in my group. We carried out investigations of photoinduced electron transfer (124), and exciton transfer (107).  Jiaji Zhang  investigated the  proton transfer and proton coupled electron transfer process in conjunction with 2D spectra (129, 135). Yuki Iwamoto calculated 2D THz spectra using rotationally invariant SB models (126). 
Calculating nonlinear spectra and comparing the results with those obtained from the HEOM, Tatsushi Ikeda demonstrated the applicability of the surface hopping method (124). 

For investigations of 2D spectroscopy, Seiji Ueno used machine learning techniques to create a SB model with ground and excited state PESs based on microscopic trajectories obtained from MD and quantum chemical calculations (127, 136). The constructed model is then used to compute various 2D vibrational and electronic spectra using the HEOM approach. This should improve our ability to perform large-scale computations. 

By calculating the nonlinear response function, we were able to identify peculiar effects of quantum dissipative dynamics which manifest as experimental observables, such as 2D spectra. Arend Dijkstra, a postdoc from the Netherlands, demonstrated this point by characterizing the role of a non-Markovian SB interaction in the quantum information process (86, 92).  Hyeon-Deuk Kim analytically investigated a mode-mode interaction in the 2D IR spectrum.

\subsection*{Quantum statistical thermodynamics}

In 2014, I worked in Freiburg Institute for Advanced Studies. I explored the foundations of the HEOM on my own and I became more convinced that this approach is sufficiently reliable for application to fundamental problems in physics (103,105).
Thus, Akihito Kato and Souichi Sakamoto, who thought that philosophy is more important than potential future employability, studied quantum thermodynamics on the basis of numerical experiments conducted using the HEOM. The distinctive feature of the HEOM approach is that it can rigorously evaluate changes in heat and entropy not only for the system but also for the bath and the SB interaction (117). We thus showed that quantum thermodynamics can be described in the framework of statistical mechanics, even in the nonequilibrium case, by regarding nonequilibrium work as a change in quasi-static free energy (131,133).

\subsection*{Solid state physics problems}
Even since my time at the UIUC, I have been interested in applying the HEOM formalism as a quantum thermostat to problems in solid-state physics. Lipeng Chen, Mauro Cainelli, and I have derived the HEOMs for various Holstein systems (109, 132). My goal is to simulate a superconducting state in real time under a time-dependent external force. At the moment, however, Kiyoto Nakamura has only been able to solve the HEOM for a four-site Holstein-Hubbard system (137). I think we will need to wait another 15 years for the necessary computational developments.

\section*{Epilogue}
Since coming to Kyoto, my international friendships have expanded. I have visited many countries for conferences, seminars, lecture courses, and sightseeing. I also have hosted many guests from abroad. I always take them to the my favorite sushi bar. Graham, Vladimir, Dwayne, and Tõnu Pullerits are such VIPs. Because of this, my fame as the ``sushi professor" has spread around the world, and the number of guests continues to grow. 

With David Coker, Yang Zhao, Maxim Gellin, Howe-Siang Tan, Frank Grossmann, and Raffaele Borrelli, I have been organizing the "Summit Meeting," a self-organized and self-assembled scientific meeting held at places where the participants particularly wish to visit. The science reported at these meetings has been of the highest level, and meetings conclude with a mountain trek. Each committee member strongly insisted on where they wished to go, so whenever the committee needs to chose a location, there is always a heated discussion.

It has been more than 40 years since I started my career in science. At first, I thought that I was climbing alone with no peak in sight. However, in fact,
many friends and students have been accompanying me on my journey---still with no peak in sight. As I keep climbing, my {\it fortune} indeed continues {\it rising}. 
Now, finally, I appreciate my name, {\it Yoshitaka}.\\

\textbf{Yoshitaka Tanimura}

\newpage
\clearpage
\section*{Publications of Yoshitaka Tanimura}
\cite{TTK86JPSJ,TK89JPSJ1,TK89JPSJ2,TSK89JPSJ3,TK89JPSJ4,T90PRA,TW91PRA,TW92JCP,TM93PRE,TM93JOSA,
TM93JCP,TM93JPC,TM94JPSJ,TM94JCP,PBSLTM94JPC,TM95JCP,OT96PRE,OT96JCP,TTT96CPL,OT97JCP1,
TO97JCP,TM97JCP,OT97JCP2,OT97PRE,OT97CPL1,OT97CPL2,COT98JCP,TTK98JCP,T98CP,GT98CPL,
OT98CPL,MT98CPL,ITYT98CPL,ITT99JPC,ST99PRE,GGT99CP,OTT99JCP,AHTS99PRB,ITYT99JCP,ITT99SM,
TYA99PNAS, OTT99CPL,OBT20BCSJ,ST20JPSJ,TS20JPSJ,OJT01CP,KT01CPL,ST01JPSJ,ST01JCP,HTT01JCP,
ST02CPL,HTT02CPL,TLO02JCP,ST02JPSJ,DT02JO,KT02JCP1,ST03JCP,OT03JCP,OT03JPC,KT04JCP,
IT05JCP,IT05JPSJ,KT05JCP,NT06JCP,ST06JCP,NHT06JCP,HT06JCP,T06JPSJ,IT06JCP,ST07JCP,
NTM07JCP,KTC07JCP,IT07JPCA,IT08CP,HT08JCP,KTC08JCP,ST08JCP,JT08CPL,LHDHT08JCP,MTH09ACR,
TI09ACR,TT09JPSJ,UTP10JPCL,DT10NJP,TT10JCP,DT10PRL,ST11JPCA,HT11JPCB,PT11PRE,UTP12JPCL,
UTP12JPCC,DT12PTRA,DT12JPSJ,DT12NJP,HSOT12JCP,PT12PRE,T12JCP,UTT12IJQC,ST13JPSJ,KT13JPCB,
GTD13JCP,ST14NJP,T14JCP,IHT14JCP,T15JCP,IIT15JCP,DT15JCP,TT15JCTC,CYT15JPCL,KT15JCP,
IJT15SD,GST16JPSJ,IT16JCP,ZCHSTZ16JPC,JIT16CP,IHT16JPCL,KT16JCP,WRTM17NJP,IT17JCP,ST17JPCL,
NT18PRA,IT18CP,IT18JCP,IT19JCTC,IDT19JCP,IT19JCP,UT20JCTC,TT20JPSJ,ZBT20JCP,T20JCP,
ST20JCP,CT21JCP,ST21JSPS,JT21JCP,ZBT21JCP,UT21JCTC,NT21JCP,IT21JCEL}

\bibliography{tanimura_publist}

\newpage
\clearpage
\section*{Colleagues of Yoshitaka Tanimura}
\subsection*{Former Research and Visiting Students}
\begin{tabbing}
 \hspace{40mm} \= \hspace{40mm}  \kill
Maruyama, Yutaka \>
Imamura, Yutaka \\
Hino, Osamu  \>
Nagata, Yuki \\
Ishizaki, Akihito \>
Suzuki, Yoichi  \\
Hasegawa, Taisuke \>
Joutsuka, Tatsuya \\
Ueta, Satoshi \>
Ono, Junichi \\
Sakurai, Atsunori \>
Tanaka, Midori \\
Ito, Hironobu \>
Kato, Akihito \\
Tsuchimoto, Masashi \>
Ikeda, Tatsushi \\
Jo, Ju-Yeon \>
Sakamoto, Souichi \\
Nakamura, Kiyoto \>
Steffen, Thomas \\
Chen, Lipeng\>
Sakumichi, Naoyuki\\
Otaki, Hiroki  \>
Yoshimune, Seiji \\
Uchikoshi, Motonobu\>
Shimada, Arisa \\
Oosawa, Yu \>
Usui, Kota \\
Nawata, Haruchika \>
Hanaoka, Yoshiki \\
Li, Baiquing \>
Schumacher, Anne\\
Lee, Chee Kong \>
Ianto,Cannon\\
Sun, Shining\>
Kong, Fanchen \\
Simbananiye, Jordan \>
\end{tabbing}

\subsection*{Former Postdoctoral Research Fellows and Research Associates}
\begin{tabbing}
 \hspace{40mm} \= \hspace{40mm} \kill
Okumura, Ko \>
Suzuki, Yoko \\
Tomita, Kenichi  \>
Gangopadhyay, Gautam   \\
Okada, Akira \>
Miyazaki, Kunimasa \\
Kato, Tsuyoshi \>
Kim, Hyeon-Deuk  \\
Kuninaka, Hiroto \>
Horikoshi, Atsushi \\
Dijkstra, Arend G. \>
Packwood, Daniel \\
\end{tabbing}
\subsection*{Present Group Members}
\begin{tabbing}
 \hspace{40mm} \= \hspace{40mm}  \kill
Ueno, Seiji \>
Iwamoto, Yuki \\
Umehara, Kazuki   \>
Zhang, Jiaji  \\
Cainelli, Mauro  \>
Liu, Zifeng  \\
Takahashi, Hideaki \>
Park, Kwanghee \\
Zhang, Yankai \>
Koyanagi, Shoki \\
Hoshino, Ryotaro \>
\end{tabbing}
\subsection*{Past and Present Collaborators}
\begin{tabbing}
 \hspace{40mm} \= \hspace{40mm}  \kill
Kubo, Ryogo   \>
Takano, Hiroshi \\
Wolynes, Peter G \>
Mukamel, Shaul   \\
Miller, R. J. Dwayne \>
Ten-no, Seiichiro L. \\
Cho, Minhaeng   \>
Tokmakoff, Andrei\\
Jonas, David M.\>
Bagchi, Biman   \\
Yonemitsu, Kenji   \>
Klafter, Joseph \\
Onuchic, Jose N \>
Leite, Vitor B. P,  \\
Khün, Oliver  \>
Du, Si-de  \\
Hirata, Fumio \>
Sethia, Ashok  \\
Ando, Koji \>
Prezhdo, Oleg V. \\
Hamm, Peter\>
Gelin, Maxim F. \\
Domcke, Wolfgang   \>
Grossmann, Frank \\
Zhao, Yang \>
Mintert, Florian \\
Witt, Bjorn \>
Rudnicki, {\L}ukas \\
Borrelli, Raffaele \>
Paquette, Glenn \\
\end{tabbing}

\section*{Abbreviated Curriculum Vitae of Yoshitaka Tanimura}
Data of Birth: June 26, 1960 

\subsection*{Education}
 \noindent
$\bullet$Keio University, Department of Instrumentation \\(with R. Kubo)    1984/3    B.S\\
$\bullet$Keio University, Department of Instrumentation \\(with R. Kubo)    1986/3    M.S.\\
$\bullet$Keio University, Department of Physics \\(with R. Kubo)            1989/3    Ph.D\\

\subsection*{Professional Experience}  
\noindent
$\bullet$Beckman Institute, University of Illinois at Urbana-Champaign  \\
Postdoctoral fellow   (with P. G. Wolynes, K. Hess, and A. J. Leggett) 1989/5-1992/9 \\
$\bullet$Department of Chemistry, University of 
Rochester\\
Post-doctoral Fellow (with S. Mukamel) 1992/10-1994/3\\
$\bullet$Institute for Molecular Science \\
Associate professor 1994/4-2003/5 \\
$\bullet$Department of Chemistry, Kyoto University \\
Professor 2003/6-present\\

\noindent
$\bullet$ School of Chemistry, Tel Aviv University,
Visiting Professor  1995/11 \\
$\bullet$ Solid State and Structural Chemistry Unit, Indian Institute of Science,
Visiting Professor  1997/1 \\
$\bullet$ Institute for Molecular Science,
Visiting Professor  2012/6-2012/7 \\
$\bullet$ Department of Chemistry, Technical University of Munich,
Visiting Professor  2010/3-2011/3  \\
$\bullet$ Department of Physics, University of Augsburg,
Visiting Professor  2012/6-2012/7 \\
$\bullet$ Department of Physics, University of Hamburg, DESY,
Visiting Professor  2012/2-2012/5\\
$\bullet$ Freiburg Institute for Advanced studies, Freiburg University,
Senior Research Fellow  2014 \\
$\bullet$ Department of Chemistry, Beijing University,
Visiting Professor  2019 \\

\subsection*{Professional Service}
\noindent
$\bullet$Journal of Physical Society of Japan\\
Editorial board member  1998/4-2002/3 \\
$\bullet$Japanese Physical Society    \\
Board member    2002/4-2004/3 \\
$\bullet$Fukui Institute for Fundamental Chemistry, Kyoto University\\       
Vice Director,    2015/4-2018/3  \\
$\bullet$Faculty of Science, Kyoto University \\ 
Vice Dean of Science,     2020/4 --\\

\subsection*{Awards}
\noindent
$\bullet$Morino Science Foundation Award, 2002 \\
$\bullet$Humboldt Research Award, 2012\\
$\bullet$Yagami Award, 2013\\
$\bullet$American Physical Society Fellow, 2015\\
\end{document}